# Coexistence of Superconductivity and ferromagnetism in high entropy carbide ceramics

Huchen Shu[1,2], Wei Zhong[1], Jiajia Feng[1], Hongyang Zhao[2], Fang Hong[3,4,5*], Binbin Yue[1*]

[1] Center for High Pressure Science & Technology Advanced Research, 10 East Xibeiwang Road, Haidian, Beijing 100094, China

[2] School of Science, Hubei Key Laboratory of Plasma Chemistry and Advanced Materials, Wuhan Institute of Technology, 206, Guanggu 1st Road, Wuhan 430205, China

[3] Beijing National Laboratory for Condensed Matter Physics, Institute of Physics, Chinese Academy of Sciences, Beijing 100190, China

[4] School of Physical Sciences, University of Chinese Academy of Sciences, Beijing 100190, China

[5] Songshan Lake Materials Laboratory, Dongguan, Guangdong 523808, China

*Email: hongfang@iphy.ac.cn; yuebb@hpstar.ac.cn

**Generally, the superconductivity was expected to be absent in magnetic systems, but this reception was disturbed by unconventional superconductors, such as cuprates, iron-based superconductors and recently discovered nickelate, since their superconductivity is proposed to be related to the electron-electron interaction mediated by the spin fluctuation. However, the coexistence of superconductivity and magnetism is still rare in conventional superconductors. In this work, we reported the coexistence of these two quantum orderings in high entropy carbide ceramics $(Mo_{0.2}Nb_{0.2}Ta_{0.2}V_{0.2}W_{0.2})C_{0.9}$, $(Ta_{0.25}Ti_{0.25}Nb_{0.25}Zr_{0.25})C$, and they are expected to be conventional superconductors. Clear magnetic hysteresis loop was observed in these high entropy carbides, indicating a ferromagnetic ground state. A sharp superconducting transition is observed in $(Mo_{0.2}Nb_{0.2}Ta_{0.2}V_{0.2}W_{0.2})C_{0.9}$ with a $T_c$ of 3.4 K and upper critical field of ~3.35 T. Meanwhile, superconductivity is suppressed to some extent and zero-resistance state disappears in $(Ta_{0.25}Ti_{0.25}Nb_{0.25}Zr_{0.25})C$, in which stronger magnetism is presented. The upper critical field of $(Ta_{0.25}Ti_{0.25}Nb_{0.25}Zr_{0.25})C$ is only ~1.5 T, though they show higher transition temperature near 5.7 K. The ferromagnetism stems from the carbon vacancies which occurs often during the high temperature synthesis process. This work not just demonstrate the observation of superconductivity in high entropy carbide ceramics, but also provide alternative exotic platform to study the correlation between superconductivity and magnetism, and is of great benefit for the design of multifunctional electronic devices.**

In 2004, the concept of high-entropy alloys (HEAs) was proposed by Yeh et al. and Cantor et al. [1, 2], which rapidly attracted a significant amount of research interest due to their highly disordered and homogeneous single-phase characteristic [3, 4]. Subsequently, Rost et al.[5] synthesized a stable face-centered cubic structured high-entropy oxide ($Mg_{0.2}Co_{0.2}Ni_{0.2}Cu_{0.2}Zn_{0.2}$)O in 2015, demonstrating for the first time the entropy stabilization of oxides and introducing the concept of high-entropy into the field of ceramics [5, 6]. Since then, new types of high-entropy materials such as borides [7], carbides [8], nitrides [9], silicides [10], fluorides [11], and hydrides [12] have continuously emerged, and their applications have been found in thermal protection [13], supercapacitors [14], wear-resistant coatings [15], biocompatible coatings [16], water splitting [17], nuclear reactor cladding [18], and so on.

As early as 1972, transition metal carbides (TMCs) were recognized for their significant solid solubility [19]. In addition, the coexistence of metallic and covalent bonds gives them excellent properties such as high melting point, admirable hardness, nice electrical/thermal conductivity, and so on [20, 21]. High-entropy carbides (HECs) are generally a simple single-phase system composed of multiple metal carbides (≥5) that are molten in equimolar proportions, and often exhibit superior performance compared to single-component systems. Studies have found that HECs display superior properties such as higher hardness [22, 23], better wear resistance [24], more excellent high-temperature stability [25], and oxidation resistance [26] than traditional monolithic TMCs. Furthermore, The superconductivity in TMCs was reported several decades ago[21]. In 1962, Giorgi et al. [27] investigated the relationship between the superconducting transition temperature ($T_c$) and carbon contents in Ta-C and Nb-C carbides, and found that the closer the C/X (X represents a metal atom) molar ratio is to 1, the higher the $T_c$. Soon after, Willens et al. [28] claimed that the reason for the influence of $T_c$ in binary TMC alloys may be related to lattice disorder scattering, based on the study of the mutual solubility in various binary TMC phases, and measured the $T_c$ of NbC, MoC, TaC, and WC, which was almost equal to previously reported data. The superconducting and topological properties of TaC and NbC were studied by Shang et al. [29] recently. The $T_c$ values are found to be 10.3 K and 11.5 K, respectively, and band structure calculations show that the density of state on the Fermi level is mainly controlled by the $d$ orbitals of Ta or Nb and strongly hybridizes with the C $p$ orbitals to form a large cylindrical Fermi surface, similar to high-$T_c$ iron-based superconductors. In the same year, Yan et al. [30] prepared single-crystal NbC using solid-phase reaction method and measured its $T_c$ as high as 12.3 K. The experiment showed that NbC belongs to type-II superconductor, judging from the behavior of upper and lower critical magnetic fields. It exhibits strong Fermi surface nesting, and this is beneficial for the strong electron-phonon interaction, which ultimately enhances the superconductivity. In addition, Ge et al. [31] measured the $T_c$ of α-$Mo_2C$ to be 7.5 K. Hence, it is reasonable to believe that HECs should possess superconductivity, if it includes one or more superconducting metal carbides. The explore of superconductivity in high entropy compounds was initiated firstly in HEA. Koželj et al. [32] discovered the superconductivity of $Ta_{34}Nb_{33}Hf_8Zr_{14}Ti_{11}$ in 2014, proving it to be a type-II superconductor with a measured $T_c$ of 7.3 K. Since then, many superconducting phenomena have been reported in HEAs [33-35], and subsequently, superconducting phenomena have also been found in high-entropy oxides [36, 37]. Recently, Liu et al. [38] reported the discovery of superconductivity in high-entropy silicides for the first time, with a relatively high $T_c$ value (3.2-3.3 K). Unfortunately, research on HECs has mainly focused on their thermodynamic properties [22-26], and there are rare studies on their superconducting properties to date.

In this work, we synthesized and characterized two types of HECs: $(Mo_{0.2}Nb_{0.2}Ta_{0.2}V_{0.2}W_{0.2})C_{0.9}$ and $(Ta_{0.25}Ti_{0.25}Nb_{0.25}Zr_{0.25})C$ (named as Mo-HEC and Ta-HEC, respectively). X-ray diffraction (XRD) analysis showed that both of them are single NaCl phase (*Fm-3m*, No.225), in which the transition metal atoms are randomly distributed on the cationic positions and the carbon atoms occupy the anionic positions. Superconductivity was observed in Mo-HEC at 3.4 K through low temperature electrical transport measurements with an upper critical field of ~3.35 T, which is below the weak-coupling Pauli limit (1.84 T/K * 3.4 K ≈ 6.26 T), and suggests a typical conventional superconductor. Meanwhile, magnetic measurement shows a ferromagnetic ground state in the whole temperature range. Previous studies demonstrate that the ferromagnetic ordering is an intrinsic behavior in carbides with metal element and carbon vacancies [39, 40]. Previous theoretical calculation claims that the *p* electrons of the nearest-neighbor carbon atoms near the vacancies is responsible for the long-range ferromagnetic ordering [39]. Hence, it will be of great interests to observe the coexistence of ferromagnetism and superconductivity. To examine the universality of such kind of phenomenon in HEC, we studied another compound Ta-HEC. A much higher $T_c$ ≈ 5.7 K is observed in Ta-HEC but zero-resistance state is absent. Magnetism measurement demonstrated a much stronger ferromagnetism existing in Ta-HEC (at least 20 times of that in Mo-HEC), which is expected to suppress the superconductivity.

Microstructure analysis shows that Ta-HEC has a smaller average grain size than Mo-HEC (~12 μm *VS* 42 μm). This means that Ta-HEC should have a higher formation energy and higher vacancy rate. Carbon content was investigated by an infrared carbon/sulfur analyzer and verified that 96.55% of ideal carbon content is found in Mo-HEC, while only 81.9% is found in Ta-HEC. Our work provides a promising way to realize the coexistence of ferromagnetism and superconductivity in HECs, which may also apply to other high-entropy superconducting system with light elements (C, B, O and N et al.).

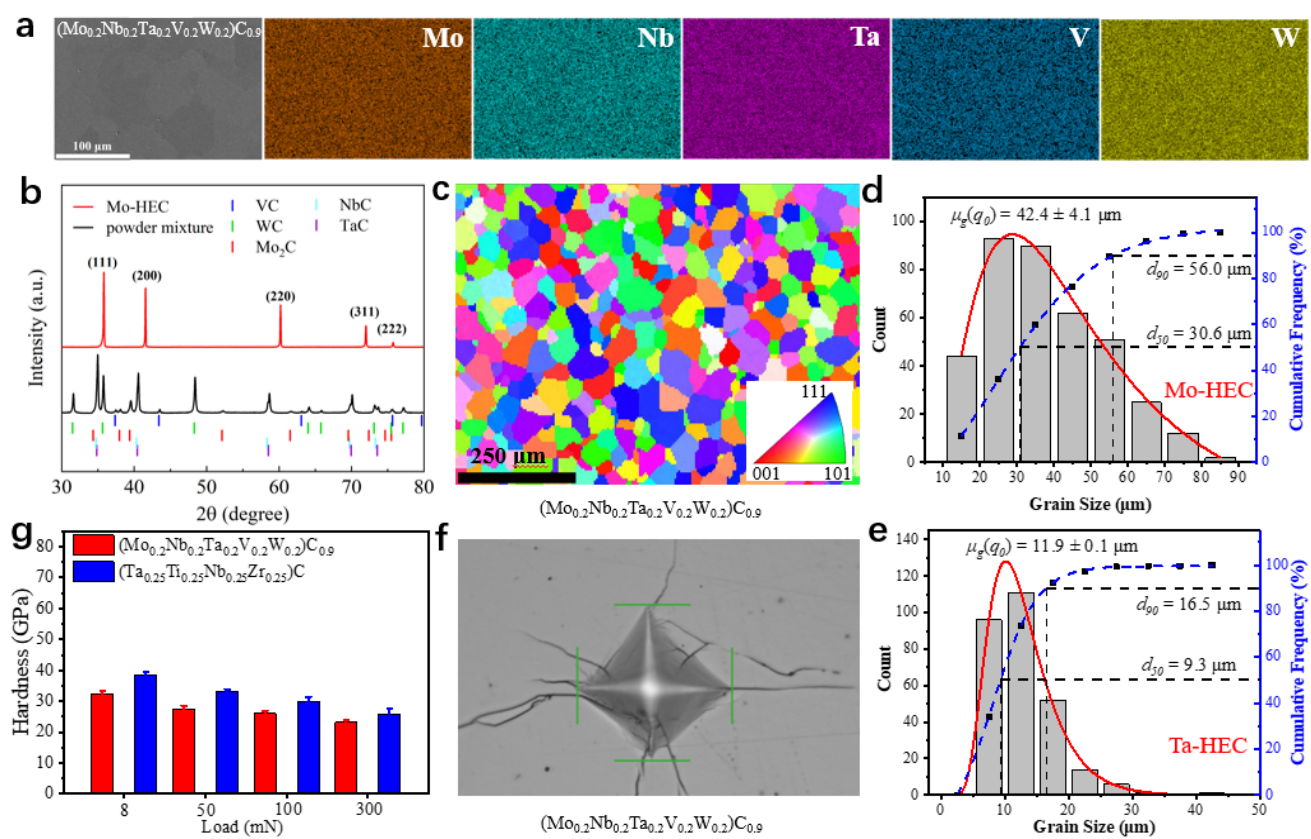

**Fig.1 The structure and morphology characterization of $(Mo_{0.2}Nb_{0.2}Ta_{0.2}V_{0.2}W_{0.2})C_{0.9}$ (Mo-HEC) and $(Ta_{0.25}Ti_{0.25}Nb_{0.25}Zr_{0.25})C$ (Ta-HEC) high entropy carbides. a.** The SEM image and corresponding individual element distribution analyzed by EDX in Mo-HEC, **b.** the X-ray diffraction pattern of sintered Mo-HEC, pristine mixture powder and original phases of various binary carbides used in this work, **c.** the EBSD image of Mo-HEC showing the grain size, grain boundaries and orientation, **d.** statistical result of grain size distribution for Mo-HEC, **e.** statistical result of grain size distribution for Ta-HEC, **f.** the Vickers-hardness testing image on a Mo-HEC, **g.** the nanoindentation results for both Mo-HEC and Ta-HEC.

Mo-HEC and Ta-HEC high entropy carbides were synthesized by sparking plasmon sintering method. The mole fraction of five (or four) metal elements is equal. Fig.1 shows the elemental analysis, morphology and structure characterization results. The EDX results show an even distribution of each principal metal, as seen in Fig.**1a**. The sintered sample is a simple cubic phase with only five diffraction peaks in our experimental range of x-ray diffraction, while the pristine powder mixture shows a complex pattern with contribution from each binary metal carbide, as seen in Fig.**1b**. The starting powder was grounded by ball milling and the grain size was only 1-2 microns or even submicron, as shown in Supplementary information, but the final product shows a large grain size, as shown in Fig.**1c** and Fig.**1d**. Statistical grain size is 42.4±4.1 microns for Mo-HEC. The grains show random orientation as seen by the Electron Backscatter Diffraction (EBSD) image in Fig.**1c**. For comparison, the grain size of Ta-HEC is only 11.9±0.1 microns. Different grain size distribution suggests that Ta-HEC has a higher formation energy than Mo-HEC. The smaller grain size also signals that there would be more grain boundaries and defects existing in Ta-HEC, and this will be discussed later. The hardness of Mo-HEC and Ta-HEC was briefly checked by a Vickers-hardness tester, and the HECs is very fragile, as seen in Fig.1**f**. Then, a delicate nanoindentation measurement was carried out, as seen in Fig.1**g**, and Mo-HEC has a hardness of ~26 GPa at 100 mN loading while Ta-HEC has a little bit higher hardness of ~ 30 GPa.

Electrical transport property of Mo-HEC was investigated by a commercial Physical Property Measurement System, equipment with a 9T magnet (PPMS-9). At high temperature range, the resistivity of Mo-HEC doesn't change too much and is almost constant below 40-50 K (Fig.2a). A sharp drop of resistivity is observed near 3.4 K, and zero-resistance state is presented as well, suggesting the occurrence of superconductivity. The magnetic field suppresses the superconductivity as $T_c$ shifts to lower temperature with pressure (Fig.2b). The zero-temperature upper critical field is extracted by fitting the upper critical field-$T_c$ relation with the Ginzburg–Landau (G-L) formula as presented in Fig.2c, which is about 3.37 T. Such an upper critical field is smaller than the Bardeen-Copper-Schrieffer (BCS) weak coupling Pauli paramagnetic limit of $\mu_0 H_P = 1.84 T_c = 6.25$ T for $T_c \approx$ 3.40 K, suggesting the absence of Pauli pair breaking.

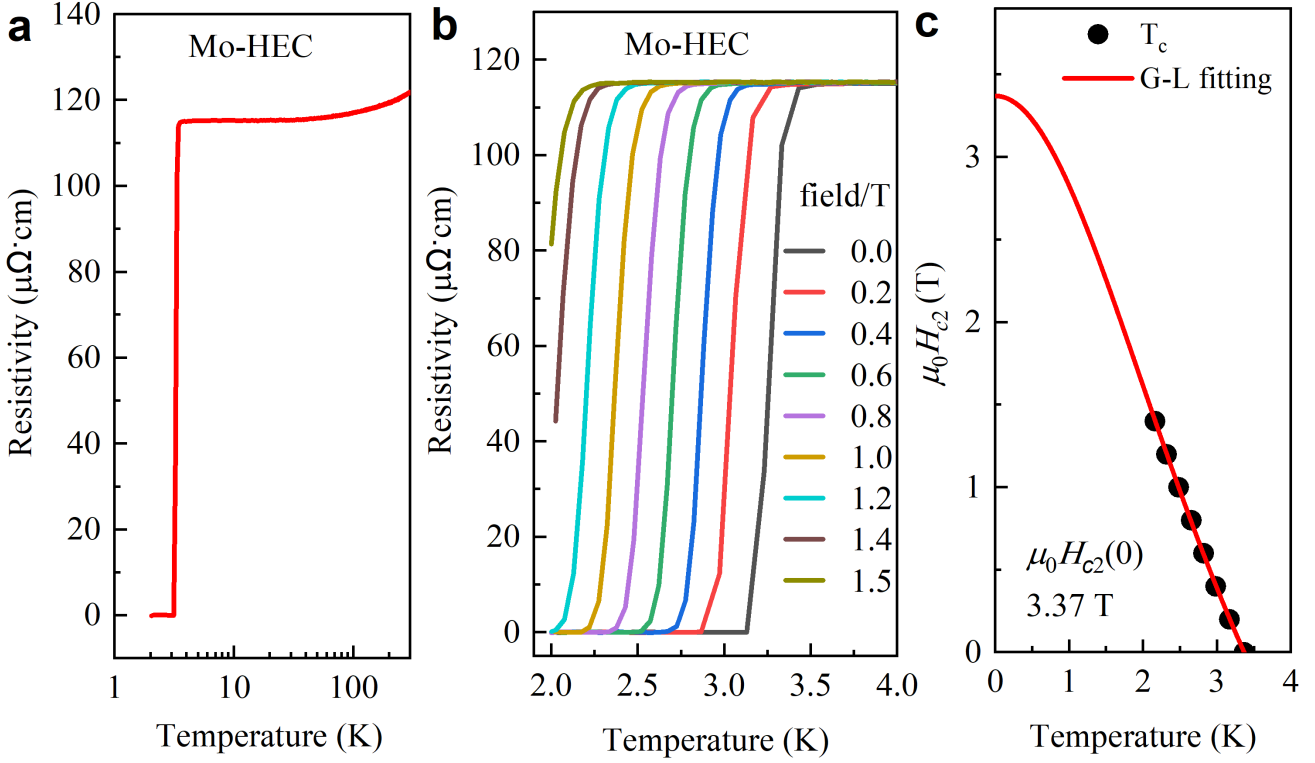

**Fig.2 The electrical transport properties of Mo-HEC. a.** The RT curve is collected from 2 K to 300 K, a sharp superconducting/SC transition is observed near 3.4 K, **b.** the magnetic field effect on the SC transition, **c.** the upper critical field-Tc relation extracted from **b**, and the curve was fitted by Ginzburg–Landau (G-L) formula, giving a zero-temperature upper critical field of ~3.37 T.

To further verify the superconductivity, magnetism measurement was carried out. As shown in Fig.3a, Mo-HEC shows a clear diamagnetic behavior below the superconductivity. It is noted that there is a small temperature deviation (~0.2 K) of $T_c$ from R-T measurement, which can be ignored since the magnetic measurement is carried out in a MPMS system, in which the temperature sensor set-up is different from that in PPMS. A typical diamagnetic hysteresis loop is also observed, as seen in Fig.3b. These magnetic results prove that there is the Meissner effect in Mo-HEC and it is truly a superconductor. Meanwhile, the magnetic behavior is also studied above SC transition, as shown in Fig.3c and Fig.3d. Mo-HEC shows a clear ferromagnetic signal at 4 K, just above the $T_c$. The ferromagnetic signal should not be from any impurity but from the carbon vacancy induced magnetism, which has been proposed in previous work on carbon-based compounds [39, 40].

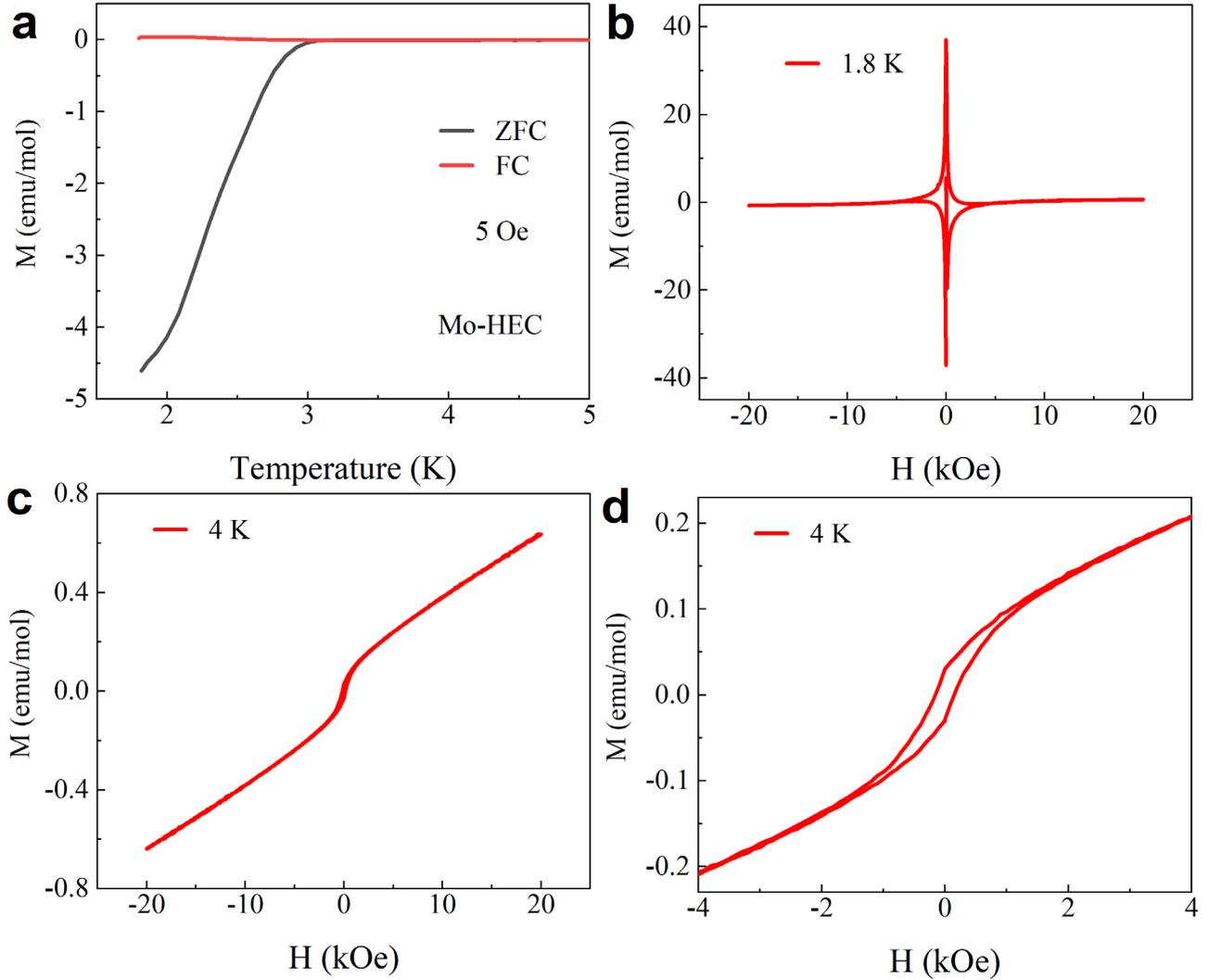

**Fig.3 The magnetic measurement of Mo-HEC high entropy carbide. a.** The ZFC-FC curves near the SC transition, showing a typical Meissner effect with clear diamagnetic signal, and it was further confirmed by the magnetic hysteresis loop at 1.8 K presented in **b**. **c.** The M-H curve measured at 4 K, which is above the SC transition temperature, and zoon-in region with a clear ferromagnetic loop is seen in **d**.

To check whether it is an accidental result showing the coexistence of superconductivity and magnetism in HEC, the other high entropy carbide-$(Ta_{0.25}Ti_{0.25}Nb_{0.25}Zr_{0.25})C$ (Ta-HEC) has also been tested. The transport results are displayed in Fig.4. The resistivity of Ta-HEC shows a little bit more sensitive temperature dependence at high temperature. Similarly, its resistivity is also almost constant below 40-50 K, while there is sharp drop near 5.7 K, signaling a possible SC transition, since zero-resistance state is absent this time (Fig.4a). Meanwhile, there is also another drop below ~4 K, suggesting a two-phase SC transition. The magnetic field effect verifies the existence of superconductivity (Fig.4b), since the $T_c$ shifts to lower temperature while the resistivity of normal metal state is a constant, excluding the possibility of a magnetoresistance behavior. The G-L fitting giving a zero-temperature upper critical field of ~1.59 T (Fig.4c), which is obviously lower than the Bardeen-Copper-Schrieffer (BCS) weak coupling Pauli paramagnetic limit.

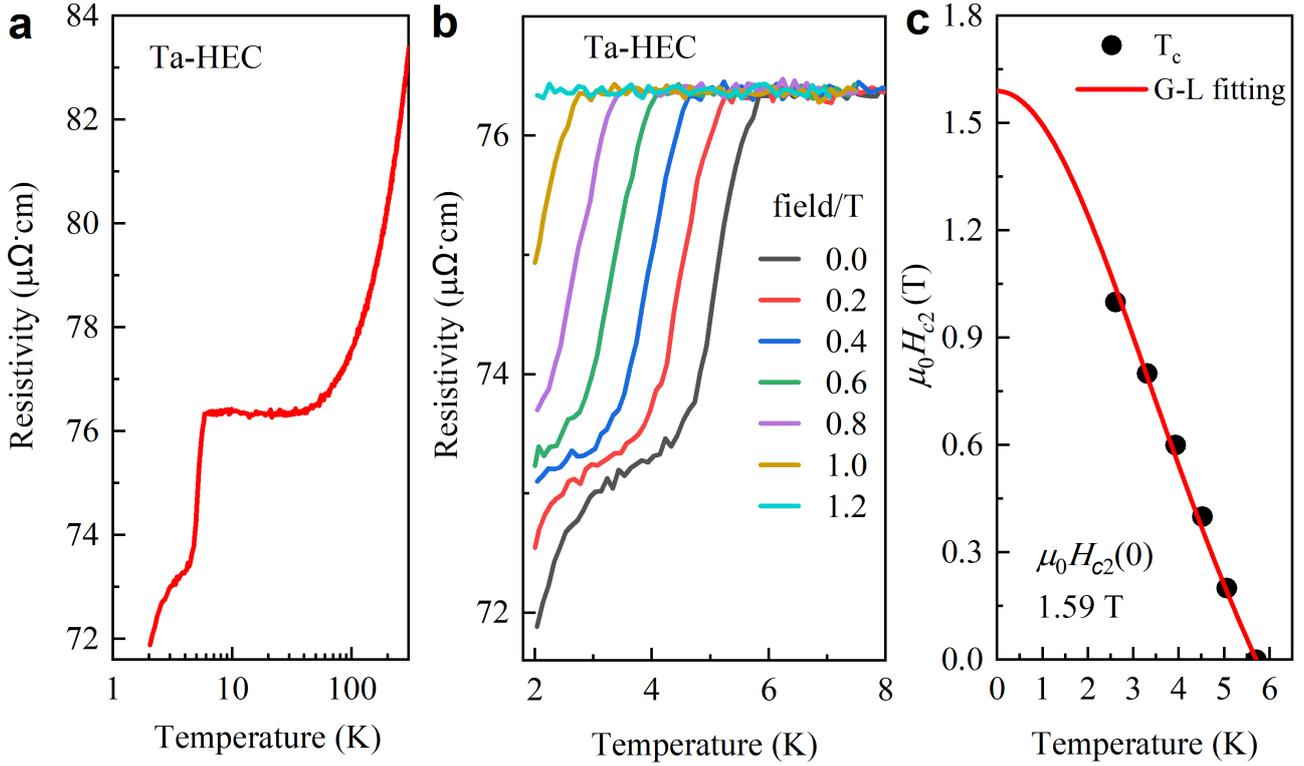

**Fig.4 The electrical transport properties of $(Ta_{0.25}Ti_{0.25}Nb_{0.25}Zr_{0.25})C$ (Ta-HEC). a.** The RT curve is collected from 2 K to 300 K, a sharp drop of resistivity is observed near 5.7 K, while another drop is near 4 K, suggesting possible superconducting/SC transitions, **b.** The magnetic field effect verifies the SC transition, **c.** the upper critical field-$T_c$ relation extracted from **b**, and the curve was fitted by Ginzburg–Landau (G-L) formula, giving a zero-temperature upper critical field of ~3.37 T for the higher SC phase starting near 5.7 K.

To further verify the superconductivity, the magnetic measurement was carried out. Fig.5 shows the ZFC-FC curve and magnetic hysteresis loop measured at various temperature. It is clear that Ta-HEC has a stronger ferromagnetic background than that in Mo-HEC. The separation of ZFC and FC signal is even observed near room temperature, as seen in Fig.5a, and a sharp drop of magnetic moment is observed at low temperature. After zooming in this region, it is found that the drop behavior in temperature dependent magnetic moment is consistent with the SC transitions in transport measurement. If the ferromagnetic background was extracted from the ZFC-FC curves, the modified ZFC-FC curves will show a typical diamagnetic behavior (not shown). Therefore, the sum effect of SC diamagnetic behavior and strong ferromagnetic background finally gives an overall positive moment rather than a negative one. The ferromagnetic background can be tracked by the magnetic hysteresis loop measured at various temperatures, as displayed in Fig.5c and Fig.5d. It is clear that there is also strong ferromagnetic signal even in the SC state. At 200 K, the FM signal is similar with that at 1.8 K. It is noted that the remanent magnetic moment in Ta-HEC is much higher than that in Mo-HEC (2 emu/mol vs 0.05 emu/mol), suggesting a strong FM ground state, which affects the superconductivity much more strongly and cause the absence of zero-resistance state in Ta-HEC.

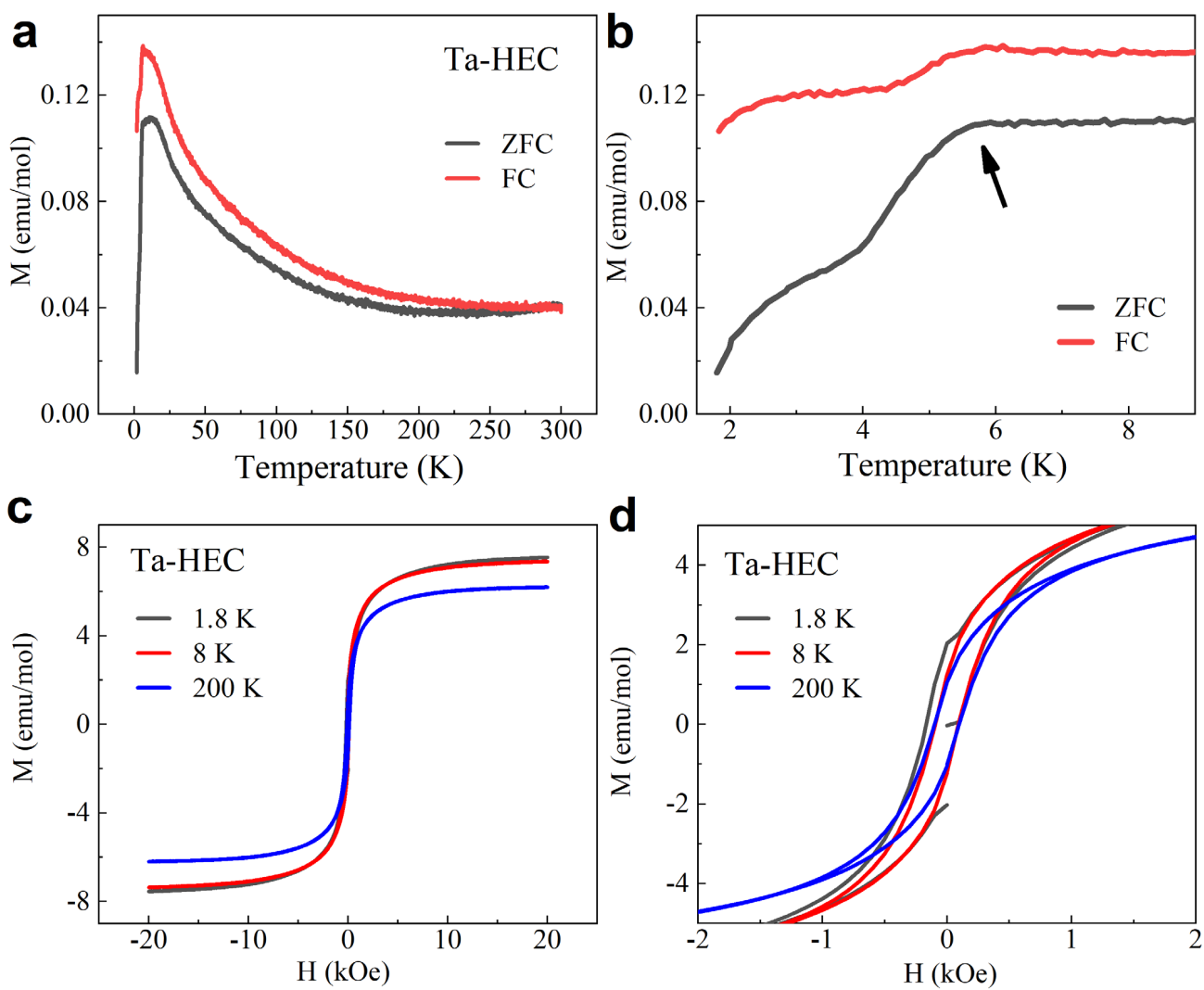

**Fig.5 The magnetic measurement of Ta-HEC. a.** The ZFC-FC curves shows a relatively strong

ferromagnetic background at high temperature range while the magnetic moment shows clear drop at low temperature near the SC transition, **b.** the zoom-in region near SC transition, a Meissner effect is presented, which is embedded in the FM background, and the magnetic measurement finally gives an overall positive moment rather than a typical diamagnetic negative moment. Two transitions can be seen and consistent with transport measurement. **c.** The M-H curve measured at 1.8 K, 8 K and 200 K respectively, which covers the SC transition region and normal metallic state as well, **d.** the zoom-in region with a clear ferromagnetic loop is seen in all temperature points, suggesting a strong FM ground state.

To verify the correlation between carbon vacancy and ferromagnetism in these two HECs, we did more analysis on the carbon content and vacancy by using an infrared carbon/sulfur analyzer (CS-800, ELTRA GmbH, Germany). As shown in Table I, the Mo-HEC has a theoretical 1:0.9 molar ratio for the metal and carbon, the experimental result gives a value of 1:0.869, which means a 96.55% carbon content, and 3.45% carbon vacancy. For Ta-HEC, the ideal molar ratio is 1:1 for the metal and carbon, however, the measured metal/carbon ratio is only 1:0.819, which means only 81.9% carbon content and 18.1% carbon vacancy. This testing result demonstrates that there are much more vacancies existing in Ta-HEC. This is also consistent with the grain size distribution analysis, and Ta-HEC has a much smaller average grain size, which means more grain boundaries and defects. Such a large difference can well explain the large ferromagnetic signal in Ta-HEC.

**Table I The carbon content and carbon vacancy analysis**

| Sample | Theoretical metal/carbon molar rate | Testing sample weight/mg | Carbon weight | Experimental metal/carbon molar rate | Carbon content (Exp/theory) | Carbon vacancy |
|---|---|---|---|---|---|---|
| Mo-HEC | 1:0.9 | 22.6 | 7.8785 % | 1:0.869 | 96.55% | 3.45% |
| Ta-HEC | 1:1 | 17.5 | 8.6964 % | 1:0.819 | 81.9% | 18.1% |

Molecular weight (g/mol): Mo-HEC, 131.73; Ta-HEC, 115.25

Therefore, it could be a prevailing phenomenon that there is coexistence of superconductivity and magnetism in HEC. The magnetism in HEC is generally a ferromagnetic behavior and does not favor the superconductivity. However, the $T_c$ values in different HECs may vary a lot and the coexistence behavior may inspire the protype design of multifunctional devices. Meanwhile, when we finished our work, we also noted that there is report about the superconductivity in HEC ceramic- $(Ti_{0.2}Zr_{0.2}Nb_{0.2}Hf_{0.2}Ta_{0.2})C$, and this HEC has a $T_c$ of ~2.5 K and upper critical field of ~0.51 T [41]. Clearly, the $T_c$ and upper critical field in our Mo-HEC and Ta-HEC is much higher. Since there is ferromagnetic background in Mo-HEC and Ta-HEC, the $T_c$ should be enhanced if the ferromagnetic behavior is suppressed by tuning the carbon vacancies and synthesis conditions, which is worthy of further study.

Such vacancy-induced magnetism may also exist in other compounds showing superconductivity, such as borides or oxides in form of simple compounds or high entropy compounds (since the light element can easy escape during high temperature synthesis process), especially for transition metal-based compounds. Meanwhile, the manipulating of vacancy and corresponding SC and/or magnetism by external methods, such as liquid ion gating, will also be of great interests, and the small ion intercalation to the vacancy position (such as $Li^+$, $H^+$) would be a promising way to achieve higher $T_c$ in high entropy ceramics.

## Methods

**1) Sample preparation**

In this study, two types of HEC were prepared: $(Mo_{0.2}Nb_{0.2}Ta_{0.2}V_{0.2}W_{0.2})C_{0.9}$ and $(Ta_{0.25}Ti_{0.25}Nb_{0.25}Zr_{0.25})C$ (The following are referred to as Mo-HEC and Ta-HEC respectively). The initially selected carbide powders were TaC (99.5%, 1-4 μm), TiC (99.99%, 1-12 μm), NbC (99%, 1-4 μm), ZrC (99%, 1 μm), $Mo_2C$ (99.95%, 1-4 μm), and WC (99.9%, 400 nm), all purchased from Shanghai Aladdin Biochemical Technology Co., Ltd., China. VC (99.9%, 1-4 μm) was purchased from Adamas-beta, Shanghai Titan Scientific Co., Ltd., China. Each single carbide precursor was weighed in equimolar amounts of metal atoms, and then the precursor powders were poured into a 50 ml stainless steel vacuum ball mill jar lined with $ZrO_2$ and configured with different diameter $ZrO_2$ grinding balls (diameter 3.15 mm-10 mm) to achieve a ball-to-powder ratio of approximately 5:1, using ethanol as the milling medium. The ball mill jar was evacuated using a mechanical pump and milled at 500 rpm for 6 h using a planetary ball mill (TJX-410, Tianjin Oriental Tianjing Technology Development Co. Ltd., China), with a 10 min pause every 0.5 h to reduce the possibility of oxide formation due to overheating during milling. After milling, the powder was dried in a 70 °C drying box for 8 h, and then siezed and placed into a graphite mold with a diameter of 10 mm, with a layer of carbon paper between the powder and the mold to facilitate demolding. The mold was placed in a spark plasma sintering furnace (SPS-3T-3-MINI (H), Shanghai Chenhua Technology Co., Ltd., China), and the sample chamber was evacuated and filled with Ar. Under the initial pre-pressure of 11.25 MPa, the temperature was set to rise from room temperature to 700 °C within 5 min, and then to 1800 °C at a rate of 100 K/min, and kept for 10 min to remove residual gas in the powder. The temperature continued to rise to 2050 °C within 3 min, while the pressure was increased to 30 MPa and kept under this condition for 3 min, and then within 4 min, the temperature was raised to 2200 °C, and the pressure was increased to 50 MPa, and maintained for 6 min. The sample was then cooled at a rate of 100 K/min to 800 °C and then to room temperature with the furnace, resulting in an HEC sample with a diameter of 10 mm and a thickness of approximately 2 mm.

**2) (Micro) structure characterization**

X-ray diffraction (XRD) measurements were performed using a PANalytical Empyream X-ray Diffractometer (Holland) with a copper target and a wavelength of $\lambda(K_\alpha)$ = 1.5406 Å, after grinding the samples. The detector is PIXcel3D, and the x-ray Cu target works on a voltage of 40 kV and a current of 40 mA. Scanning was conducted in the range of 5°-80° with a step size of 0.013° within 8 mins.

Microstructural analysis and element distribution uniformity were investigated using a scanning electron microscope (SEM, JSM-7900F of JEOL, Japan) equipped with an Oxford X-Max N50 Aztec EDS detector at 15 kV through secondary electron scattering. Due to the large error of the EDS detector in the determination of the content of light elements such as C, we chose to use an infrared carbon/sulfur analyzer (CS-800, ELTRA GmbH, Germany) to measure the carbon content of the HECs. The detection is based on GB/ T 20123-2006, the working carrier gas is oxygen, compressed air is used as the power gas, and the detection limit is 0.1 ppm.

Furthermore, crystal grain size distribution and orientation determination were carried out using SEM (VERSA 3D, FEI, Hillsboro, OR, USA) equipped with an electron backscatter diffraction (EBSD) detector at 20 kV. Raw EBSD data were analyzed and post-processed using the Tango module in HKL Channel 5 software. Mo-HEC and Ta-HEC were characterized with scanning step sizes of 4 μm and 1.4 μm, respectively.

3) **Electrical transport and magnetic property measurement**

A piece of stick-like sample with well-defined geometry was cut from the as-prepared HEC sample by using a diamond sawing system. Then, the electrical transport measurements from 2 K to 300 K were carried out on a commercial Physical Property Measurement System (PPMS, Quantum Design), based on a standard four-probe method.

The magnetic measurements were carried out on a Vibrating Sample Magnetometer (VSM) incorporating with the PPMS. A small magnetic field 5 Oe is used to study the temperature dependent magnetism during the zero-field cooling and field-cooling, which helps to study the Meissner effect and verify the superconductivity. Meanwhile, the magnetic hysteresis loops were measured at representative temperatures, showing the ferromagnetic behavior.

4) **Mechanical property testing**

The Vickers hardness $H_V$ and fracture toughness $K_{IC}$ were measured using a micro- Vickers hardness tester (Qness 60A$^+$, Germany) equipped with a Vickers diamond indenter under a load of 9.8 N for 15 s. The $H_V$ and $K_{IC}$ were determined by the following equations [39]:

$$H_V = \frac{1854.4F}{L^2} \qquad (1)$$

$$K_{IC} = \frac{0.016(E/H_V)^{0.5} F}{C^{1.5}} \qquad (2)$$

here $F$ (N) is the applied load, $L$ (μm) is the arithmetic mean diagonal length of the indentation, $C$ (μm) is the average length of the radial cracks and $E$ is the Young's modulus of HEC.

Nanoindentation tests were performed using a nanoindentation system (KLA G200, Milpitas, CA,

USA) equipped with a standard Berkovich diamond indenter. Prior to testing, the Berkovich indenter was calibrated using fused silica. In the nanoindentation test, four sets of maximum load tests of 8 mN, 50 mN, 100 mN and 300 mN were selected for each sample, the loading rate was 0.5 mN/s, and the maximum load was stayed for 5 s. The thermal drift was kept below 0.05 nm/s and corrected using the measured value at 10% of the full load during unloading. Using Continuous Stiffness Measurement (CSM) mode, using a 5 × 5 indentation array for indentation mapping under each load, with a 20 μm distance between adjacent indentations to ensure that it is not affected by the residual stress field of adjacent indentations, and grain measurement data for several different orientations were collected. SEM (VERSA 3D, FEI, Hillsboro, OR, USA) observations were performed on the collected indentation arrays, and indentation data at grain boundaries or defects (such as pores) were eliminated to make the results more consistent. The Young's modulus ($E$) was obtained from the loading/unloading-displacement curves according to the Oliver-Pharr method [40]. The values of Vickers hardness, nanoindentation hardness and Young's modulus ($E$) were the average values of at least five measurements.

## Acknowledgment

This work was supported by the National Key R&D Program of China (Grants No. 2021YFA1400300), the Major Program of National Natural Science Foundation of China (Grants No. 22090041), the National Natural Science Foundation of China (Grants No. 12004014, No. U1930401). Part of the experimental work was carried out at the Synergic Extreme Conditions User Facility.